\documentclass[groupedaddress, showpacs, superscriptaddress]{revtex4-2}
\usepackage{amssymb,amsmath,amsthm,color,graphicx,times,graphicx}
\usepackage{hyperref}
\usepackage{subfigure}
\usepackage{times}
\usepackage{bbold}
\usepackage{color}
\usepackage{array}
\usepackage{sidecap}
\usepackage[french]{babel}
\usepackage{array,multirow,makecell}
\usepackage[table]{xcolor}

\theoremstyle{plain}

\theoremstyle{definition}

\begin{document}
\title{Controlled quantum teleportation between discrete and continuous physical systems}
\author{M. El Kirdi}\affiliation{LPHE-Modeling and Simulation, Faculty of Sciences, Mohammed V University in Rabat, Rabat, Morocco.}
\author{A. Slaoui}\email{abdallah.slaoui@um5s.net.ma}\affiliation{LPHE-Modeling and Simulation, Faculty of Sciences, Mohammed V University in Rabat, Rabat, Morocco.}\affiliation{Centre of Physics and Mathematics, CPM, Faculty of Sciences, Mohammed V University in Rabat, Rabat, Morocco.}
\author{N. Ikken}\affiliation{LPHE-Modeling and Simulation, Faculty of Sciences, Mohammed V University in Rabat, Rabat, Morocco.}
\author{M. Daoud}\affiliation{Department of Physics, Faculty of Sciences, University Ibn Tofail, Kenitra, Morocco.}\affiliation{Abdus Salam International Centre for Theoretical Physics (ICTP), Strada Costiera, 11 I-34151, Trieste, Italy.}
\author{R. Ahl Laamara}\affiliation{LPHE-Modeling and Simulation, Faculty of Sciences, Mohammed V University in Rabat, Rabat, Morocco.}\affiliation{Centre of Physics and Mathematics, CPM, Faculty of Sciences, Mohammed V University in Rabat, Rabat, Morocco.}

\begin{abstract}
Quantum teleportation of an unknown state basing on the interaction between discrete-valued states (DV) and continuous-valued states (CV) presented a particular challenge in quantum technologies. Here we consider the problem of controlled quantum teleportation of an amplitude-matched CV qubit, encoded by a coherent state of a varied phase as a superposition of the vacuum- and single-photon optical states among two distant partners Alice and Bob, with the consent of controller, Charlie. To achieve this task, we use an hybrid tripartite entangled state (interaction between the discrete and continuous variables states) as the quantum resource where the coherent part belongs to Alice, while the single-photon belongs to Bob and Charlie and the CV qubit is at the disposal of Alice. The discrete-continuous interaction is realized on highly transmissive beam-splitter. We have shown that the perfectly of teleportation fidelity depends on the phase difference between the phase of the state to teleport and the phase of the sender's mode, we found that for a difference which approaches 0 or $\pi$, near perfect controlled quantum teleportation can be obtained in terms of the fidelity and independently of the amplitude $\alpha$ and the squeezing parameter $\zeta$. Experimentally, this proposed scheme has been implemented using linear optical components such as beam splitter, phase shifters and photon counters.
\par
\vspace{0.25cm}
\end{abstract}
\date{\today}

\maketitle
\section{ Introduction}

In 1993, Bennet and his collaborators \cite{Bennett1993} suggested a physical process for transferring quantum information from one location say, Alice, to another location some distance away, Bob, by encoding information in a quantum state using a bipartite entangled resource. This protocol was successfully implemented experimentally in 1997 \cite{Boschi1998}, allowing the particles to be transferred between the two locations. This called standard quantum teleportation (SQT). Actually, there are three classes where a quantum teleportation can be classified, one via the discrete-variable (DV) entangled resource channel \cite{Bennett1993,Marshall2014}, two by the continuous-variable (CV) entangled resource channel \cite{Vaidman1994,Braunstein1998} and three via the coupling discrete-continuous variables entangled resource state \cite{Brask2010,Lim2016,Ulanov2017,Podoshvedov2019}. In 1998, that was extended to controlled quantum teleportation (CQT) involving the third location say, Charlie as a supervisor for the protocol process using GHZ state by Karlsson et al \cite{Karlsson1998}. Furthermore, in Ref.\cite{Dong2011} the authors performed a controlled communication between the three locations using GHZ state and imperfect Bell state measurement (BSM). After the sender makes BSM and the supervisor performs photon counters, the receiver carries out unitary transformation on his own particle to obtain the original state from the sender and under the cooperation of the supervisor \cite{Xiao-Ming2007}.\par

Subsequently, several theorists and experimenters have discussed, analysed and reported on this kind of quantum communication. For the later, Ting et al.,\cite{Ting2005} proposed a controlled and secure quantum teleportation of a 2-level particle via an initially shared triplet of entangled particles under the control of the supervisor Charlie. Chen et al.,\cite{Chen2010} suggested a quantum teleportation protocol between multiple senders and multiple receivers via only one supervisor Charlie, where the supervisor shares the entangled state with every sender, while there is no directly shared entanglement between sender and receiver. Wang et al.,\cite{Wang2011} presented a scheme for CQT with Bell states in which classical keys for supervisor's portion are used. Then, a new scheme to execute bidirectional CQT by using five-qubit entangled state as quantum channel \cite{Li2013,Chen2014}. Verma et al.,\cite{Verma2016} proposed precise and simple CQT of an arbitrary $N$-qubit information state using general quantum channel and measurement bases. After, a new scheme of bidirectional CQT via a five-qutrit entangled state was proposed \cite{Ma2018}. Moreover, Barasi\`nski et al.,\cite{Barasinski2018} analysed the CQT protocol throught three-qubit mixed states, investigating the relation between the faithfulness of CQT and the entanglement of shared resource state. Later, an experimental implementation of tripartite CQT on quantum optical devises was reported \cite{Barasinski2019}.\par

The principle idea of CQT is that teleportation of the encoded information cannot be complemented successfully without  the intervention and permission of the third location Charlie as supervisor, which can be exploited perfectly to communicating a quantum secret information among several locations of controllers. Since CQT takes its effective role in quantum networks and cryptography \cite{Li2008,Na2015}, a problem arises when the controller is not honest. In this case, the information security is really in danger and Bob cannot receive the original target or at least not the correct one. The theoretical idea was further extended experimentally in teleportation of discrete as well as continuous variables \cite{Vaidman1994,Kim2001}.\par

On the other hand, hybrid systems arise when we combine continuous variables that obtains values in a continuum and a discrete variables that takes values in a countable set \cite{Alur1995}. This remains a little about the same in the classic treatment of information where both types of encoding used are known as digital and analogue processing of information. Actually, most systems in our real-world are described in a hybrid way. Obviously, systems needs to be a coupling between discrete and continuous variables. Recently, several research has been made to fill both types of achieving protocols that will allow to overcome the individual limits presented by each type. Hence, hybrid systems have become a research topic that allows quantum technologies to propose protocols with more best solutions than CV or DV protocols \cite{Andersen2015}. The mixture of these two approaches (i.e., CV and DV) in one same system has found a theoretical and experimental progress. Due to this advantage, hybrid teleportation has been identified as one of the three key priorities for quantum science \cite{Pirandola2016}. Making a step towards this goal, Lvovsky and his team have develop a quantum teleportation technique by means of a hybrid entangled channel \cite{Ulanov2017}. The current paper is an extension of this contribution in which a controlled quantum teleportation using an entangled hybrid resource is proved.\par

In this paper, we will present a protocol of controlled quantum teleportation in which there are three contributors Alice, Bob and the supervisor Charlie who are going to share a quantum channel with three modes in order to transmit an unknown quantum state. This protocol cannot be carried out without the equal collaboration of these three contributors. More precisely, we show a scheme to teleport a CV qubit via a tripartite entangled resource state, encoded as a superposition of a coherent state onto a two DV modes (vacuum- and single-photon Fock states). The remaining sections of the paper were presented as follows. In Sec.II, we introduce the controlled quantum teleportation scheme with a mention of the quantum channel resource used, the state to be teleport and some necessary preliminaries for the CQT study. In Sec.III, we present the controlled quantum teleportation of an arbitrary CV-matched qubit state via an hybrid entangled state (CV-DV coupling). A concluding remarks are given in Sec.IV.
\section {Proposed controlled quantum teleportation scheme}
In Ref. \cite{Ulanov2017}, Lvovsky and his co-workers proposed a quantum teleportation scheme in which a bipartite hybrid entangled state is used. Inspired by this work, here we present a controlled quantum teleportation protocol in which a tripartite hybrid entangled state initially shared by the sender Alice, the receiver Bob, and the supervisor Charlie operates as a channel resource. The sender, the receiver and the supervisor are spatially separated from each other.
\begin{center}
	\begin{figure}[h!]
		\centering
		\includegraphics[scale=0.65]{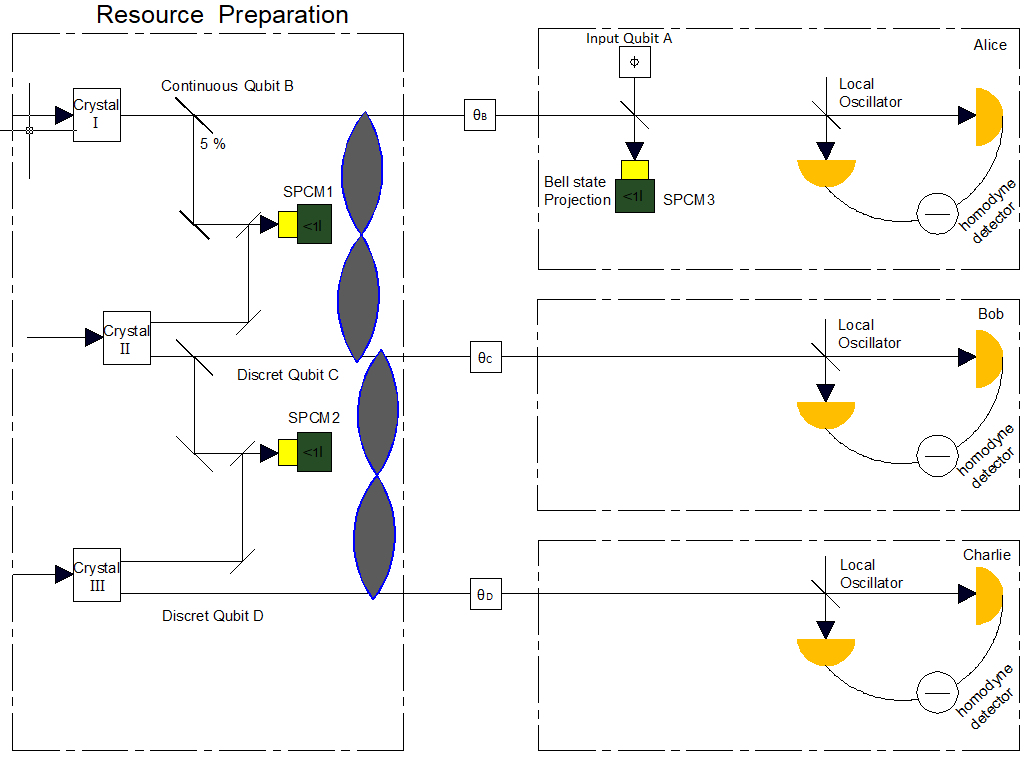}
		\caption{Process scheme that can be implemented experimentally. CQT of a continuous mode (Eq.\ref{Eq4}) via an hybrid tripartite entangled resource state (Eq.\ref{Eq1}). The resource state can be prepared by count of two single-photon counting  modules (SPCMs). It depends on when a click event of SPCM3 occurs as for Bell state measurement \cite{Feizpour2015}. After Alice's and Charlie's measurements, Bob complete the CQT using an homodyne detector in mode $C$. Two others homodyne detectors in modes $B$ and $D$ helps to continue phase measures.}\label{Fig1}
	\end{figure}
\end{center}

\subsection{ Quantum channel resource state}
Since we are interested in using a hybrid quantum resource to perform quantum teleportation, we will consider a tripartite resource state described as follows
\begin{align}
	\left|R\right>_{BCD} &= \dfrac{1}{\sqrt{2}}\left(  \left|\alpha\right>_{B}\left|0\right>_{C} \dfrac{\left|0\right>_{D}+\left|1\right>_{D}}{\sqrt{2}}
	-\left|- \alpha\right>_{B}\left|1\right>_{C} \dfrac{\left|0\right>_{D}-\left|1\right>_{D}}{\sqrt{2}}\right) \nonumber\\
	& = \dfrac{1}{2}\left(\left[\left|\alpha\right>_{B}\left|0\right>_{C}- \left|-\alpha\right>_{B}\left|1\right>_{C}\right]\left|0\right>_{D}+\left[\left|\alpha\right>_{B}\left|0\right>_{C}+\left|-\alpha\right>_{B}\left|1\right>_{C}\right]\left|1\right>_{D}\right),	\label{Eq1}
\end{align}
where $B$, $C$ and $D$ goes to Alice's, Bob's and Charlie's mode of channel resource, respectively.\par
\begin{center}
	\begin{figure}[h]
		\centering
		\includegraphics[scale=0.7]{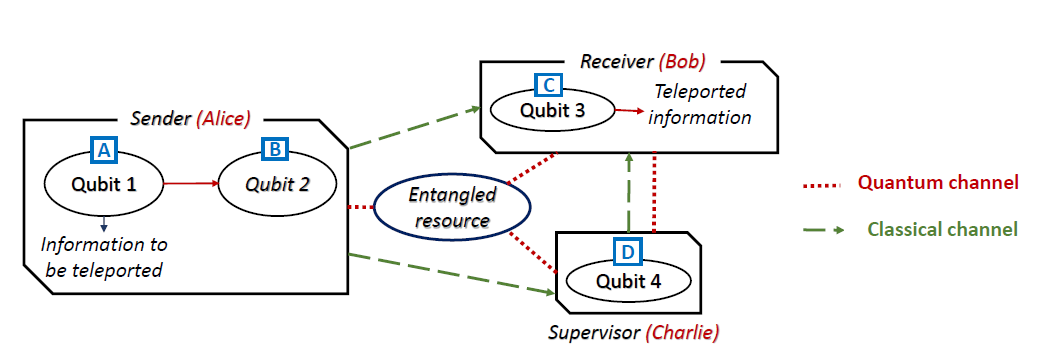}
		\caption{Conceptual scheme of the controlled quantum teleportation process.}\label{Fig2}
	\end{figure}
\end{center}

The hybrid entangled state consists of the coherent components with opposite in sign amplitudes (here and in the following, the amplitude is assumed to be positive; $\alpha>0$) and the single photon taking simultaneously two modes (vacuum and single-photon Fock states). The state (\ref{Eq1}) is non-maximally entangled state due to the non-orthogonality of the coherent states.
After the preparation of the resource of the intricate quantum channel, Alice, Bob and Charlie are now ready to start the protocol of quantum teleportation and make the list of successive tasks shown in Fig.(\ref{Fig2}). The three users Alice, Bob and Charlie are separated in the space and entangled via the quantum channel, which means that quantum information may be transmitted using the shared quantum channel between users and another channel of classical communication. The steps of the controlled quantum teleportation scheme are summarized as follows:
\begin{enumerate}
	\item After combining the state to teleport with the quantum channel, Alice carries a Bell measurement on the two modes in its possession.
	\item Then, Charlie perform a Photon-Counter (PC) measurement on his mode.
	\item To help Bob to recover information, Alice and Charlie will send to Bob the results of their measurements by classic communication channel.
	\item Based on the received results, Bob applies the appropriate unitary transformation, bit-flip ($\sigma_x$), phase-flip ($\sigma_z$) or bit-phase flip ($\sigma_y$), to acquire the initial state to teleport. 
\end{enumerate}
\subsection{State to teleport: Target state}
When one consider entangled hybrid states, directly we think about coherent states \cite{Glauber1963,Monroe1996,Zhang1990,Bakraoui2022}. Generally, coherent state are non-orthogonal, in fact the overlap for two coherent state $\left|\alpha\right>$ and $\left|\beta\right>$ is
\begin{equation}
|\left<\alpha|\beta\right>|^2 = e^{-|\alpha-\beta|^2}=x^4,
\end{equation}
where $x=\sqrt{\left<\alpha|\beta\right>}$. For orthogonal states, one can use even and odd coherent state, where
\begin{align}
	\left |ODD, \alpha \right>= N_-\left( \left |\alpha\right> -\left |-\alpha\right>\right),\hspace{1cm} \left|EVEN, \alpha\right>= N_+\left(\left|\alpha\right> + \left|-\alpha\right>\right),	\label{Eq3}
\end{align}
known as Schrödinger cat states \cite{Dodonov1974}.	Suppose that Alice has a CV qubit encoded in superposition of coherent states $\left|\alpha\right>$ and $\left|-\alpha\right>$, which she wishes to teleport to Bob. So, the input state is given by	
\begin{align}
	\left	|in\right>_A= X \left|\alpha\right> + Y \left|-\alpha\right>,	
\end{align}
whose normalization needs $|X|^2+|Y|^2+2x^2 Re\left(X^*Y \right)=1$. Using Even and Odd coherent state as in Eq.(\ref{Eq3}), the input state takes the form
\begin{align}
	\left|in\right>_A= A_+\left |EVEN, \alpha\right>_A + A_- \left|ODD,-\alpha\right>_{A},	\label{Eq4}
\end{align}
where
\begin{align}
	A_{\pm} = \dfrac{X  + Y }{2 N_{\pm}},\hspace{2cm}{\rm and}\hspace{2cm}N_{\pm} = 1/\sqrt{2(1 \pm x^2)}.	
\end{align}
\subsection{System combined state}

From Eq.(\ref{Eq1}) and Eq.(\ref{Eq4}), the combined state of the system consisting of modes $A$, $B$, $C$ and $D$ is expressed as
\begin{align}
	\left	|\psi\right>_{ABCD}& = \left|in\right>_A \otimes\left|R\right>_{BCD}\\
	& =	\dfrac{A_{+}}{2} \left|EVEN, \alpha\right>_{A}\left(\left[ \left |\alpha\right>_{B}\left|0\right>_{C}- \left|-\alpha\right>_{B}\left|1\right>_{C}\right]\left|0\right>_{D} 
	+\left[\left|\alpha\right>_{B}\left|0\right>_{C}+ \left|-\alpha\right>_{B}\left|1\right>_{C}\right]\left|1\right>_{D}\right)
	\nonumber\\
	&+\dfrac{A_{-}}{2}\left |ODD, \alpha\right>_A
	\left(\left[ \left|\alpha\right>_{B}\left|0\right>_{C}- \left|-\alpha\right>_{B}\left|1\right>_{C}\right]\left|0\right>_{D} +\left[ \left|\alpha\right>_{B}\left|0\right>_{C}+\left|-\alpha\right>_{B}\left|1\right>_{C}\right]\left|1\right>_{D}\right),\label{Eq8}
\end{align}
with the modes $A$ and $B$ are belongs to Alice. She performs a measurement on the first two modes of the combined state Eq.(\ref{Eq8}) in the quasi-Bell basis which is defined by
\begin{equation}
	\left	|\phi^{\pm}\right>_{AB} = \sqrt{2}N_+N_-\{\left|\alpha\right>_{A}\left|\alpha\right>_{B}\pm\left|-\alpha\right>_{A}\left|-\alpha\right>_{B}\},
\end{equation}
and
\begin{equation}
	\left	|\psi^{\pm}\right>_{AB}= \sqrt{2}N_+N_-\{\left|\alpha\right>_{A}\left|-\alpha\right>_{B}\pm\left|-\alpha\right>_{A}\left|\alpha\right>_{B}\}.
\end{equation}
To implement teleportation, we will discuss the projection onto each quasi-Bell state that is detected in the modes $A$ and $B$. From this measurement, we obtain
\begin{align}
	\left|out\right>_{CD}=\left<\phi^{\pm}\right|_{AB} \otimes {\left|\psi\right>_{ABCD}},
\end{align}
or
\begin{align}
	\left|out\right>_{CD}= \left<\psi^{\pm}\right|_{AB} \otimes {\left|\psi\right>_{ABCD}}.
\end{align}
\begin{table}[h!]
	\centering
	\renewcommand{\arraystretch}{1.3}
	\begin{tabular}{|c| c| c|c|}
		\hline \rowcolor{lightgray} N°  & QBM result (AB) & Photon-counter result (D) & Unitary operation \\
		
		\hline  1   & $\phi^+$ &  0   & $\widehat{U}_1$    \\
		\hline  2   & $\phi^+$ &  1   & $\openone$    \\
		\hline 3   & $\phi^-$ &  0   &   $\openone$  \\
		\hline  4   & $\phi^-$ &  1   & $\widehat{U}_1$     \\
		\hline  5   & $\psi^+$ &  0   & $-\widehat{U}_3$   \\
		\hline  6   & $\psi^+$ &  1   & $\widehat{U}_2$    \\
		\hline  7   & $\psi^-$ &  0   & $-\widehat{U}_2$    \\
		\hline  8   & $\psi^-$ &  1   & $\widehat{U}_3$    \\
		\hline
	\end{tabular}
	\caption{Applied unitary transformations after receiving Alice's and Charlie's measurement result.}
	\label{Tab1}
\end{table}
Alice then communicates her QBM result to Bob using classical channel (see Fig.\ref{Fig2}). However, even after receiving Alice's QBM result, Bob cannot yet recover the target state. At this moment, comes the supervisor's Charlie necessary role. To complete the required teleportation, Charlie should apply a photon-counter measurement on mode $D$ in $\{\left|0\right\rangle,\left|1\right\rangle\}$ basis (see Tab.\ref{Tab1}). One get two possible cases of photon-counter result for each Alice's QBM result. Out of $8$ cases, Bob tries to discover the original state by applying a suitable unitary operation $\widehat{U}$ or $\openone$ (Pauli's matrices or identity operator) for every cases of eight. For cases $1$ and $4$, Bob applies $\widehat{U}_1=\sigma_z$ and $I$ for cases $2$ and $3$. For cases $8$ and $5$, he applies $\widehat{U}_3=i \sigma_y$ and $-\widehat{U}_3$ respectively, and $\widehat{U}_2=i \sigma_x$ and $-\widehat{U}_2$ for cases $6$ and $7$.\par

Applying unitary transformations $\widehat{I}$, $\widehat{U}_1$, $\widehat{U}_2$ and $\widehat{U}_3$, Bob found the same final state 
\begin{align}
	\left|out\right\rangle_{C} = X\left|0\right\rangle_{C} + Y\left|1 \right\rangle_{C},\label{eq13}
\end{align}
where $X$ and $Y$ are the coefficients of the target state described by Eq.(\ref{Eq4}). Similar to Ref.\cite{Ulanov2017}, we replaced the CV qubit by a coherent state of a varied phase as
\begin{align}
	\left|in\right\rangle_{A}& = \left|\alpha e^{i \phi}\right\rangle_{A}\notag\\
	&= \dfrac{1}{\sqrt{2}}\left(\left|0\right\rangle_{A} +e^{i \phi}\left|1\right\rangle_{A}\right)\notag\\
	&=\dfrac{1+e^{i \phi}}{{2}}\left|\alpha\right\rangle_{A} + \dfrac{1-e^{i \phi}}{{2}}\left|-\alpha\right\rangle_{A},\label{Eq14}
\end{align}
which is the decomposition up to the single photon term in the Fock basis. So, one can replace
\begin{align}
	X = \dfrac{1+e^{i \phi}}{{2}},  \hspace{1cm}{\rm and}\hspace{1cm} Y = \dfrac{1-e^{i \phi}}{{2}}. 
\end{align}
By substituting these expressions in Eq.(\ref{eq13}), we get
\begin{align}
	\left|out\right\rangle_{C} = \dfrac{1+e^{i \phi}}{{2}}\left|0\right\rangle_{C} + \dfrac{1-e^{i \phi}}{{2}}\left|1\right\rangle_{C}.
\end{align} 
The efficiency of the controlled quantum teleportation process is quantified by the quantum fidelity between the input state $\left|in\right\rangle$ and the output state $\left|out \right\rangle$ and has been defined as \cite{Bennett1993,Mendonca2008,Kirdi2022,Raginsky2001}
\begin{equation}
{\cal F}=|\left<out|in\right>|^{2}.
\end{equation}
Note here that the teleported information is subject to some distortion after having been transmitted to some extent when $0<{\cal F}<1$. The teleported information is destroyed during the transmission process if the output and input states are orthogonal, so in this situation, the fidelity is close to zero. Eventually, the input state is similar to the output state if the fidelity approaches unity. In an explicit form, the measure of how our output state closes to the input state reads as ${\cal F}= \cos^{2}\left(\phi\right)/2$, which depends on the phase $\phi$ of the input state. For $\left|in\right>_{A}=\left|\pm \alpha\right>$, i.e., $\phi=\left\lbrace0,2\pi\right\rbrace$, the fidelity returns to its maximum value, otherwise it is reduced. Adopting this form of coherent state, one have $\left|\pm\alpha\right\rangle=\dfrac{1}{\sqrt{2}}(\left|0\right>\pm\left|1\right>)$ and Eq.(\ref{Eq1}) becomes
\begin{align}
	\left|R\right>_{BCD} 
	& = \dfrac{1}{2\sqrt{2}}\left(\left|000\right>_{BCD}+\left|001\right>_{BCD}-\left|010\right>_{BCD}+\left|011\right>_{BCD}+\left|100\right>_{BCD}+\left|101\right>_{BCD}+\left|110\right>_{BCD}-\left|111\right>_{BCD}\right).	\label{eq11}
\end{align}	
\subsection{Wigner representation and affect of phase shifts on homodyne detectors}

For continuous variables in which quantum information is encoded in infinite spectrum systems, the eigenstates $\left|x\right\rangle$ of a continuous-valued operator $\widehat{q}$ are used alternatively to the standard finite-level encoding. This operator and its conjugate $\widehat{p}$ are presented by the quadratures amplitude $q$ and $p$ \cite{Kato2005}. In general, the information contained in an arbitrary quantum state $\left|\psi\right\rangle$ is encoded in a quasi-distribution probability known as the Wigner function ${\cal W}\left(q,p\right)$. Among the sampling continuous variables, we can mention the coherent states as a type of the Gaussian states, and the Schrödinger cat states as a type of the non-Gaussian states, which have simple Wigner functions. In this system, the measurement is performed with a continuous projector, which is implemented with homodyne detection for one optical field \cite{Ludvigsen1998}. In quantum information, the Wigner quasi-probability distribution of a quantum state is used to clarify the existence of quantum entanglement \cite{Wigner1997,Hillery1984}. Indeed, the negativity of this Wigner function presents an important sign of non-classicality which reflects the presence of quantum interference in the phase space \cite{Kenfack2004,Szabo1996}. However, this negativity is not the only sign of quantumness.	
\subsubsection{Wigner function of Gaussian state}
Gaussian states are characterized by their ease of appearance, manipulation and protection from environmental effects \cite{Werner2001}. This gives them an advantage to quantum optics experimenters \cite{Walls2007}. The Wigner function of such states is always positive and is said to have a Gaussian Wigner function. Many Gaussian states have been used in quantum optics experiments such as; ($i$) quantum vacuum state (Gaussian Fock state) \cite{Davies1996}, ($ii$) coherent states (or quasi-classical states) \cite{Zhang1990} and ($iii$) squeezed states \cite{Wu1986}. To implement our controlled quantum teleportation process schematized in Fig.(\ref{Fig1}), we restrict our focus to coherent states as a continuous physical state of Gaussian type. Indeed, the Wigner representation of a coherent state $\left|\alpha\right>$ has the same Wigner representation as a quantum vacuum state with a displacement along the $q$ and $p$ axis of the phase plane with respect to the amplitude value $\alpha$. It can be expressed as \cite{Botelho2019}
\begin{equation}
	{\cal W}_{\alpha}\left(q,p\right)= \dfrac{1}{\pi}\exp\left[\dfrac{-(q-q_0)^2}{2}\right]\exp\left[{-2(p-p_0)^2}\right]
\end{equation} 
where $(q,p)$ is the pair position-momentum, $q_0$ and $p_0$ are respectively $Re(\alpha)$ and $Im(\alpha)$ of the setting $\alpha$. 
\subsubsection{Wigner function for non-Gaussian states}
Typically, Gaussian states are all characterized by the positivity of the Wigner function and their fluctuations can be interpreted as statistical noise. According to the Hudson piquet's theorem \cite{Hudson1974}, there is a link between the Gaussian character and the positivity of the Wigner function. Indeed, a pure state whose Wigner function values are all positive is a Gaussian state. The purely quantum characterization will be highlighted when we leave the Gaussian sector where the Wigner functions are all positive for a sector where the Wigner functions also take negative values, which corresponds to the non-Gaussian domain \cite{Sorenson1971}. The latter includes two specific states; the Fock states (except for the vacuum state) and the Schrödinger cat states \cite{Monroe1996}. In the following paragraph, we will focus on the second particular state.\par

To find the Wigner function of a Schrödinger cat state, we must be placed in the case where $\alpha$ is a real number, which means that the origin of the phase plan is chosen ($Im(\alpha)= 0$). The general structure of the Wigner function for a Schrödinger cat state is \cite{Kenfack2004}
\begin{equation}
	{\cal W}_{\left|\psi\right>}\left(q,p\right) = {\cal W}_{+\alpha}\left(q,p\right) + {\cal W}_{-\alpha}\left(q,p\right) + {\cal W}_{int}\left(q,p\right),
\end{equation} 
with
\begin{equation}
	{\cal W}_{\pm \alpha}\left(q,p\right) = \dfrac{N^2}{2\pi}\left[ \exp\left[{-(q\pm q_0)^2}{-2(p-p_0)^2}\right]\right],\nonumber
\end{equation} 
\begin{equation}
	{\cal W}_{int}\left(q,p\right)= \dfrac{N^{2}}{\pi}\left[ \cos(2pq_0)\exp\left[{-q^2}{-(p-p_0)^2}\right]\right],\nonumber
\end{equation} 
and
\begin{equation}
	N^2 = 1/\left[1+\cos\left(2p_{0}q_{0}\right)\exp{\left(-q_{0}^{2}\right)}\right].\nonumber
\end{equation} 
where ${\cal W}_{\pm\alpha}(q,p)$ is the Wigner function of a coherent state with amplitude $\pm\alpha$ and ${\cal W}_{int}(q,p)$ is the interference term, indicating quantum coherence that exist between the two Gaussian coherent states $\left|\alpha\right>$ and $\left|-\alpha\right>$. It also indicates the non-Gaussian character of the Schrödinger cat states. For a cat state $\left|\alpha\right> + \left|-\alpha\right>$, one has an even number of photons from which the notation $\left|EVEN,\alpha\right> \propto \left|\alpha\right> + \left|-\alpha\right>$ and for a cat state $\left|\alpha\right> - \left|-\alpha\right>$, one has an odd number of photons from which the notation $\left|ODD,\alpha\right> \propto \left|\alpha\right> - \left|-\alpha\right>$. Therefore, the Wigner function of a Schrödinger cat includes two Gaussian states centred in $\alpha$ and $-\alpha$ which corresponds to the statistical mixture $\rho_s=\dfrac{1}{2}\left(\left|\alpha\right>\left<\alpha\right| + \left|-\alpha\right>\left<-\alpha\right|\right)$ and a signal of oscillations representing the quantum correlations between the Gaussian states, taking values twice as higher than those giving by the two Gaussian (negative value at the origin of the phase space). Note that a cat state coherence is lost as soon as a photon is lost in the environment. This loss transforms an EVEN cat to an ODD cat and vice versa and so the interference term disappear (Schrödinger cat states weakness).
\begin{center}
	\begin{figure}[h!]
		\centering
		\includegraphics[scale=0.55]{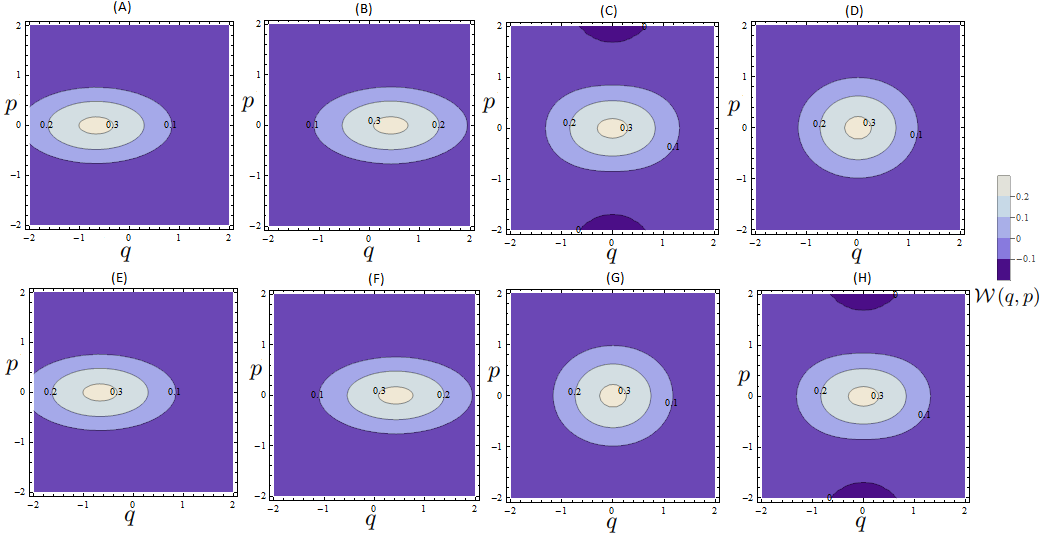}
		\caption{Contour plots of the Wigner function of the continuous mode of the quantum resource described by Eq.(\ref{Eq1}); projected in the discrete mode onto $\left|0\right>_C$ which is a coherent state (panels A and E), $\left|1\right>_C$ is a coherent state too (panels B and F) and ${(\left|0\right>_C\pm\left|1\right>_C)}/{\sqrt{2}}$, which are Schrödinger cat states (panels C, D, G and H). In upper row, Charlie's projections onto $\left|0\right>_D$ and in lower row, Charlie's projections onto $\left|1\right>_D$.}\label{Fig3}
	\end{figure}
\end{center}
In Fig.(\ref{Fig3}), we present the Wigner quasi-probability distribution of the reconstructed state, the upper row projected onto $\left|0\right>_D$ in the discrete mode $D$, first and second projected onto $\left|0\right>_C$ and $\left|1\right>_C$ in the discrete mode $C$, respectively, which are coherent states, third projected onto $(\left|0\right>_C+\left|1\right>_C)/\sqrt{2}$, which is a negative Schrödinger cat state $|ODD,\alpha>$ of amplitude $\alpha_{-}=0.73$ and fourth projected onto $(\left|0\right>_C-\left|1\right>_C)/\sqrt{2}$, which is a positive Schrödinger cat state $|EVEN,\alpha>$ of amplitude $\alpha_{+}=0.42$. In the lower row  projected onto $\left|1\right>_D$ in the discrete mode $D$, projected onto $\left|0\right>_C$ and $\left|1\right>_C$ in the discrete mode $C$, respectively, gave same results as projection in mode $D$ onto $\left|0\right>_D$ when projecting onto $\left|0\right>_C$ and $\left|1\right>_C$, projected onto $(\left|0\right>_C+\left|1\right>_C)/\sqrt{2}$ (third),  which is a positive Schrödinger cat state $|EVEN,\alpha>$ of amplitude $\alpha_+$ and projected onto $(\left|0\right>_C-\left|1\right>_C)/\sqrt{2}$ (fourth), which is a negative Schrödinger cat state $|ODD,\alpha>$ of amplitude $\alpha_-$ (here we adopt a squeezing parameter $\zeta=0.18$ \cite{Ulanov2017}).
\subsubsection{Non-local correlated measures and quadrature observables}

In quantum mechanics, the predicted result of a quantum measurement does not require the use of individual measurements but only their statistical average which is a repetitive quantity. This predicted quantity is referred to as an average value \cite{Messiah2014}. The average value of an observable $X$ in an arbitrary quantum state $\left| \psi\right> = \sum_{i=1}^{N}C_{i} \left| u_i\right>$ is defined as
\begin{align}
	\left< X \right>_{\psi} = \dfrac{1}{N} \sum_{n=1}^{N} x_n N(x_n)=  \sum_{n=1}^{N} x_n P(x_n),\label{eq23}
\end{align}
which represents the average of the results obtained by performing a number $N$ of measures on systems all in the same state, with 
$P(x_n) = |\left<u_n|\psi\right>|^2$ is the probability of measuring the own value $x_n$. The last equation (\ref{eq23}) can be rewritten as
\begin{align}
	\left< X \right>_{\psi} = \sum_{n=1}^{N} x_n \left<\psi|X|\psi\right>.
\end{align}
Here, To find out how the phase shifts of the entangled tripartite resource ($\theta_B$,$\theta_C$ and $\theta_D$) affect the correlated quadrature measured by Alice's, Bob's and Charlie's homodyne detectors\cite{Ulanov2017}, we apply the phase shift operator $\widehat{U}_{\theta_B\theta_C\theta_D} = \exp\left[{i \widehat{n}_{B}\theta_{B} + i\widehat{n}_{C}\theta_{C}+ i\widehat{n}_{D}\theta_{D}}\right]$ to the state of tripartite entangled resource (Eq.(\ref{Eq1})), which gives
\begin{align}
	\widehat{U}_{\theta_B\theta_C\theta_D}\left|R\right>_{BCD}=\left|\alpha e^{i \theta_B}\right>_B\left|0\right>_C\dfrac{\left|0\right>_D +e^{i\theta_D}\left|1\right>_D}{2}-\left|-\alpha e^{i \theta_B}\right>_Be^{i \theta_C}\left|1\right>_C\dfrac{\left|0\right>_D -e^{i\theta_D}\left|1\right>_D}{2}.
\end{align} 
Taking the expression for position quadrature observable $\widehat{X} = \left(\widehat{a}+\widehat{a}^{\dagger})\right/\sqrt{2}$ \cite{Lvovsky2015}, the predicted average value of observable $\widehat{X}_B\widehat{X}_C\widehat{X}_D$ is given by
\begin{align}
	\left\langle \widehat{X}_B\widehat{X}_C\widehat{X}_D\right\rangle=	& _{BCD}\left<R\right| \widehat{U}^{\dagger}_{\theta_B\theta_C\theta_D}          \widehat{X}_B\widehat{X}_C\widehat{X}_D\widehat{U}_{\theta_B\theta_C\theta_D}\left|R\right>_{BCD}\nonumber\\
	&= {\alpha}\sin(\theta_B-\theta_C)\sin(\theta_D).\label{eq26}
\end{align} 
where $\sin(\theta_B-\theta_C)\sin(\theta_D)$ is the probability distribution of the measurement outcome $\alpha $. Therefore, sender's mode phase  $\theta_B$, receiver's mode phase $\theta_C$ and supervisor's mode phase $\theta_D$ of the resource tripartite state can be evaluated using Eq.(\ref{eq26}), while the input coherent state phase $\phi$ is evaluated from the average reading of the sender's homodyne detector $\left<X_{\phi}\right>$. The phases of the discrete resource state, i.e., $\theta_C$ and $\theta_D$, are varied by means of the piezo-electric transducer, while the phase $\theta_B$ remains constant without activating stabilization.

\section{Controlled quantum teleportation of a coherent state}
Now, we can start the execution of our controlled quantum teleportation protocol schematized in Figure (\ref{Fig1}) on a coherent state of varied phase (eq.(\ref{Eq14})) taking into account the other three phases of the tripartite entangled resource (eq.(\ref{Eq1})). In fact, the combined state of the four modes $A$, $B$, $C$ and $D$ takes the following form
\begin{align}
	\left|\alpha e^{i \phi}\right>_A\otimes\left[\widehat{U}_{\theta_B\theta_C\theta_D}\left|R\right>_{BCD}\right] & =\left|\alpha e^{i \phi}\right>_A\left|\alpha e^{i \theta_B}\right>_{B}\left|0\right>_{C} \dfrac{\left|0\right>_{D}+e^{i \theta_D}\left|1\right>_{D}}{2}\nonumber\\
	&-\left|\alpha e^{i \phi}\right>_A\left|- \alpha e^{i \theta_B}\right>_{B}e^{i \theta_C}\left|1\right>_{C} \dfrac{\left|0\right>_{D}-e^{i \theta_D}\left|1\right>_{D}}{2},\label{eq25}
\end{align}
where the modes $A$ and $B$ at Alice's range pass through a symmetrical beam splitter, therefore the state (\ref{eq25}) becomes
\begin{align}
	\left|\psi\right>_{ABCD}=&
	\left|\alpha \dfrac{e^{i \phi}+e^{i \theta_B}}{\sqrt{2}}\right>_A \left|\alpha \dfrac{e^{i \phi}-e^{i \theta_B}}{\sqrt{2}}\right>_{B}
	\left|0\right>_{C} \dfrac{\left|0\right>_{D}+e^{i \theta_D}\left|1\right>_{D}}{{2}}\nonumber\\
	&- \left|\alpha \dfrac{e^{i \phi}-e^{i \theta_B}}{\sqrt{2}}\right>_A\left| \alpha \dfrac{e^{i \phi}+e^{i \theta_B}}{\sqrt{2}}\right>_{B}
	e^{i \theta_C}\left|1\right>_{C} \dfrac{\left|0\right>_{D}-e^{i \theta_D}\left|1\right>_{D}}{{2}}.
\end{align}
This state is then submit to a single photon detection in modes $A$ and $B$ by applying a positive operator valued measure $\widehat{\Pi}_{A}\otimes \openone_{B}$, where
\begin{equation}
	\widehat{\Pi}_{A} =\Sigma_{n}\left[1-(1-\eta)^{n}\right]\left|n\right>\left<n\right|,
\end{equation}
and $\openone_{B}$ is the identity operator (here, the measurement does not change the state of mode $B$). Hence, the bipartite output state of modes $C$ and $D$ (receiver and supervisor) takes the following form
\begin{equation}
	\widehat{\rho}_{CD} = Tr_{AB}\left[\left(\widehat{\Pi}_A \otimes \openone_{B}\right)\left|\psi_{ABCD}\right>\left<\psi _{ABCD}\right|\right].
\end{equation}
In the first order of $\alpha$, POVM is approximatively reduced to
\begin{equation} 
	\widehat{\Pi}_{A} \otimes\openone_{B}=\left|10\right>_{AB},
\end{equation}
and the output state of this projection well be 
\begin{align} 
	_{AB}\left<10|\psi_{ABCD}\right>=\dfrac{\alpha}{2\sqrt{2}}\left[\left(e^{i \phi}+e^{i \theta_B} \right) \left|0\right>_{C} \left( \left|0\right>_{D}+e^{i \theta_D}\left|1\right>_{D}\right)
	-\left(e^{i \phi}-e^{i \theta_{B}} \right)e^{i \theta_{C}}\left|1\right>_{C}\left(\left|0\right>_{D}-e^{i \theta_D}\left|1\right>_{D}\right)\right].
\end{align}
The particularity of this protocol is that the communication between Alice and Bob depends on the agreement of the third collaborator Charlie. So at this moment, Charlie must intervene and apply a photon counter measure on the mode $D$ (see the Table.(\ref{Tab2})).
\begin{table}[h!]
	\centering
	\renewcommand{\arraystretch}{1.3}
	\begin{tabular}{|p{2cm}||p{3.2cm}||p{3cm}|}
		\hline \rowcolor{lightgray} N°   & Photon counter result (D) & Unitary operation\\
		\hline  1   &  0   & $\widehat{U}_{1}$    \\
		\hline  2   & 1   & $\openone$    \\
		\hline
	\end{tabular}
	\caption{Applied unitary transformations after receiving Alice's and Charlie's measurement result.}\label{Tab2}
\end{table}

The teleport state obtained after applied Bob's unitary transformation takes this final form
\begin{align}
\left|out\right\rangle_{C}=\dfrac{1}{2}\left[\left(e^{i\phi}+e^{i\theta_B} \right) \left|0\right\rangle_{C} +  \left(e^{i\phi}-e^{i\theta_B} \right)e^{i \theta_C}\left|1\right\rangle_{C}\right],\label{eq31}
\end{align}
which define the ideal expectation. In that case, the fidelity of quantum teleportation is given by
\begin{align}
	{\cal F}=\dfrac{1}{4}\left[2+\cos\left(\theta_{B}-\theta_{c}\right)-\cos\left(2\phi-\theta_{B}-\theta_{c}\right)\right].\label{Eq32}
\end{align}
In the classical limit ${\cal F}=1/2$, which attained for $\theta_{B}-\theta_{c}=2\phi-\theta_{B}-\theta_{c}$, i.e., at $\phi=\theta_B$ then $\widehat{\rho}_{C} = \left|0\right\rangle_{CC}\left\langle0\right|$,
this means that Bob only get half the state to teleport. 
Now, fixing the receiver's mode phase $\theta_{c}$ of Bob and assuming that $\theta_{B}=\theta_{c}$, one get ${\cal F}=\left(3-\cos\left(2\left(\phi-\theta_{B} \right) \right)\right)/4$.

\begin{figure}[h!]
	{{\begin{minipage}[b]{.3\linewidth}
				\centering
				\includegraphics[scale=0.5]{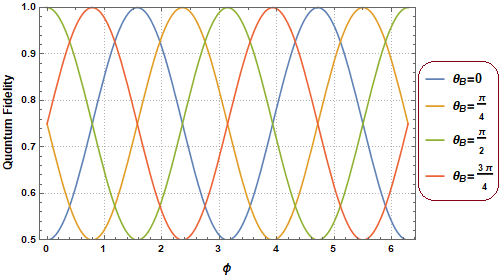}\vfill
				$(a)$
			\end{minipage}\hfill
			\begin{minipage}[b]{.24\linewidth}
				\centering
				\includegraphics[scale=0.5]{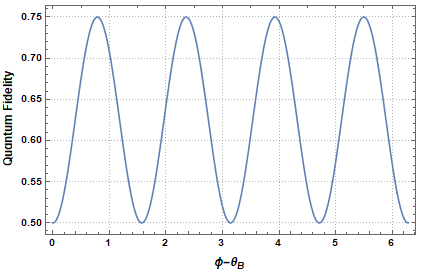}\vfill
				$(b)$
	\end{minipage}\hfill
\begin{minipage}[b]{.31\linewidth}
	\centering
	\includegraphics[scale=0.5]{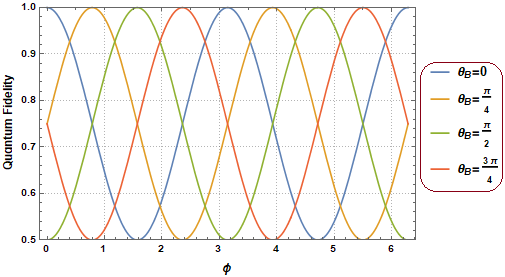}\vfill
	$(c)$
\end{minipage}}}
	\caption{($a$) Fidelity of first order of $\alpha$ after Alice's projection onto $\left|10\right>_{AB}$ (zero- and one-photon detection) and Charlie's photon-counting, obtained for $\theta_{B}=\theta_{c}$. ($b$) The same as in figure (4.$a$) but for $\theta_{B}-\theta_{C}=\pi/2$. ($c$) Fidelity of CQT in the second ordre of $\alpha$ relative to the ideal expectation.}\label{Fig4}
\end{figure}
In Fig.\ref{Fig4}($a$), we depict the variation of the fidelity of the concrolled quantum teleportation in the first order of $\alpha$ with respect to the phase $\phi$ of the teleported coherent state for various value of sender's mode phase $\theta_{B}$. In this level, one have
\begin{equation}
	\rho_{C} = \dfrac{1}{2}\begin{pmatrix}
		1+\cos\left(\phi-\theta_{B}\right)&x^{*}\left(1-y\right)\\
		x\left(1-y^{*}\right)&1-\cos\left(\phi-\theta_{B}\right)\label{eq33}
	\end{pmatrix},
\end{equation}
with
\begin{equation}
	x= \exp\left[i\phi\right]/2,  \hspace{1cm}{\rm and}\hspace{1cm}
	y= \exp\left[2i\left(\phi-\theta_{B}\right)\right]/2.
\end{equation}
Ideally, controlled quantum teleportation is perfect and teleportation efficiency (\ref{Eq32}) reach its maximum value when $\theta_{B}=\theta_{c}$ with $\phi-\theta_{B} = \pi/2$. In such case, equation (\ref{eq33}) reduces to the density matrix of the state to be teleported (\ref{Eq14}), which is exactly the original state $\rho_{C}=\left|in\right\rangle_{AA}\left\langle in\right|$.

\begin{center}
	\begin{figure}[h]
		\centering
		\includegraphics[scale=0.5]{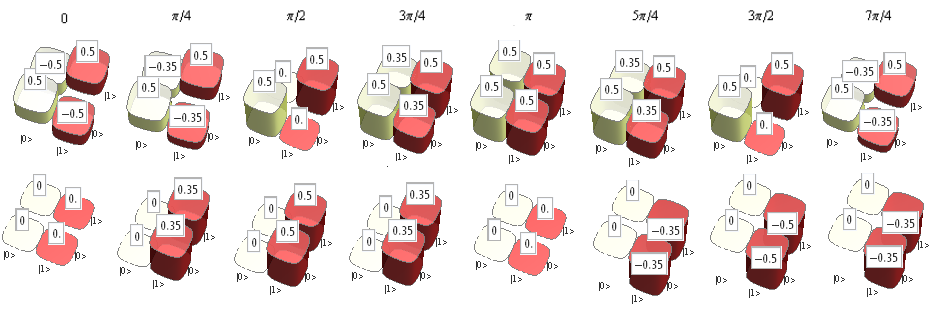}
		\caption{Density matrices of the teleport state, as measured by Bob's homodyne detector and Charlie's homodyne detector, for various phases of the input coherent state, obtained for $\theta_B-\theta_c=0$ and $ \phi-\theta_B =\pi/2$. The top row represents the real part, and the bottom row is the imaginary part.}\label{Fig5}
	\end{figure}
\end{center}	
Fig.(\ref{Fig5}) shows different density matrices of the teleportation state given by equation (\ref{eq33}), obtained for $\theta_B-\theta_c=0$ and $ \phi-\theta_B =\pi/2$ for various values of the input coherent state phase $\phi$. As provided by equation (\ref{eq11}), the diagonal elements of the teleport density matrices are independent of $\phi$ and remain equal to $1/2$ all the time, while the off-diagonal elements change their values for each $\phi$'s value between $0$ and $2\pi$, giving us an oscillatory behavior. Moreover, if $\theta_{B}-\theta_{c}=\pi/2$ were the proposed phase difference between the $B$ and $C$ modes (see Eq.(\ref{eq26})), so that the probability of $\alpha$ occurrence is perfect, the variation of the fidelity is between two values $1/2$ and $3/4$ as shown in Fig.\ref{Fig4}.($b$).\par

For the second order of $\alpha$ (one- and two-photon detection) and in the small efficiency limit $\eta<<1$, the resulting bipartite state is a mixed state. Taking the form below
\begin{equation}
	\rho_{CD} = \dfrac{1}{4}\begin{pmatrix}
		a&a f^*&-b e^*&be^*f^*\\
		a f&a&-b e^*f&be^* \\
		-ce&-cef^*&d&-df^* \\
		cef&ce&-df&d
	\end{pmatrix},
\end{equation}
with the entries given by
\begin{align}
	a =\alpha^{2}\left(1+\cos\left[\phi-\theta_{B}\right]\right),
\end{align}
\begin{align}
	b &=\dfrac{\alpha^2}{2}e^{-i(\phi-\theta_B)}\left[1-e^{2i(\phi-\theta_B)}\right]
	\left[1+\alpha^{2}(1+\cos[\phi-\theta_B])\right],
\end{align}
\begin{align}
	c &=\dfrac{\alpha^2}{2}e^{i(\phi-\theta_B)}\left[1-e^{-2i(\phi-\theta_B)}\right]
	\left[1+\alpha^{2}\left( 1+\cos[\phi-\theta_B]\right)\right],
\end{align}
\begin{align}
	d =\alpha^2\left(1-\cos[\phi-\theta_{B}]\right),\hspace{1cm}e =e^{i\theta_C} \hspace{1cm}{\rm and} \hspace{1cm}
	f =e^{i\theta_D},
\end{align}
which represent the density matrix of the two modes $C$ and $D$. The state received by Bob is
\begin{align}
	\rho_{C} &= {\rm Tr}_{D}\left[\rho_{CD}\right]=\dfrac{\alpha^2}{2}\begin{pmatrix}
		1+\cos\left(\phi-\theta_B\right)&0\\
		0&1-\cos\left(\phi-\theta_{B}\right)
	\end{pmatrix}.
\end{align}
Relative to equation (\ref{eq31}), the state (\ref{eq33}) has the quantum fidelity
\begin{align}
	{\cal F}=\dfrac{1+\cos^{2}\left[\phi-\theta_{B}\right]}{2},
\end{align}
which is independent of squeezing parameter $\zeta$ and coherent amplitude $\alpha$ of the input coherent state and reach the perfect value for $\phi-\theta_{B}=k \pi$ with $(k=0,1,2)$. \par

The variation of the quantum fidelity of the controlled quantum teleportation in the second order of $\alpha$ is depicted in Fig.\ref{Fig4}.($c$). It is observed that the minimum value of the quantum fidelity is ${\cal F}_{\min}=1/2$ and its maximum value is ${\cal F}_{\max}=1$, and these depends on both the sender mode phase $\theta_{B}$ and the target state phase $\phi$. This means that the perfect quantum teleportation can be done if we control these two parameters. In other words, for some particular and well-selected values of these two mode phases, we can say that the protocol is perfectly executed. For example, if we choose to annihilate the sender's mode phase, i.e., $\theta_B=0$, then we have two choices; either annihilate also the target state phase $\phi$, or take $\phi=2\pi$.
\section{Exeperimental realization of the proposed CQT}
In this section, we aim to present the implementation of the CQT protocol presented in this paper, where we want to teleport a coherent state of a varied phase Eq.(14), from a sender Alice to a remote receiver Bob with the help of third party Charlie, using an hybrid entangled tripartite quantum channel and classical communication. For this reason, we must first build the quantum channel pre-shared between the three parties Alice, Bob and Charlie. Here, the quantum channel was created in Alice's part, implementing several required quantum gates and operations including deferred measurements. After performing all necessary operations, Alice shares the prepared three qubits with the other responsible parties, and will immediately destroys her qubits and no longer have the ability to make any further communication.

\subsection{Preparation of the tripartite hybrid quantum channel}
\begin{figure}[h!]
	\centering
	\includegraphics[scale=0.3]{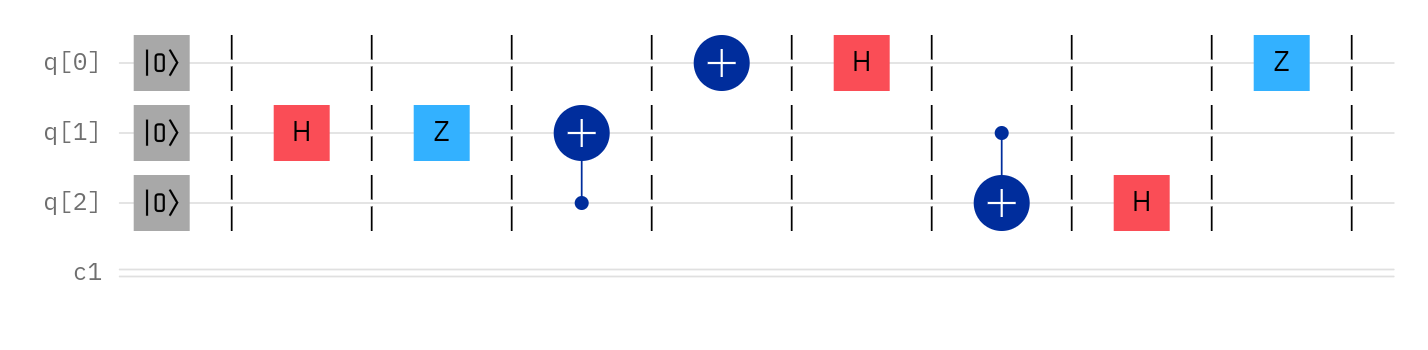}
	\caption{Quantum circuit illustrating the construction of the tripartite quantum channel. The barriers on the circuit shows the 9 steps necessary for the preparation.}\label{Fig6}
\end{figure} 
In the simplest case, where the phase of the coherent state is equal $0$ or $2\pi$, the quantum channel Eq.(1) can take the form of Eq.(18). Let us first consider this form, and after we will add several phases. The quantum channel consider here, is a three-qubit entangled state shared by the sender Alice, the receiver Bob and the supervisor Charlie, which can be built by Hadamard (H-gates), Controlled-Not (CNOT-gates) and by realizing Pauli (X,Y,Z) gates in "IBM-qasm simulator".\\
Firstly, the three-qubit entangled state is represented in the quantum circuit as product of three-qubit states all initialized to $\left|0\right>$,
\begin{align}
	\left|R_0\right>_{BCD} = \left|0\right>_{B} \otimes \left|0\right>_{C} \otimes \left|0\right>_{D},
\end{align}
where qubit B, qubit C and qubit D belongs respectively to Alice, Bob and Charlie. Then, performing H-gate on qubit C, the initial state $\left|R_0\right>_{BCD}$ is converted to 
\begin{equation}
	\left|R_1\right>_{BCD} = \dfrac{1}{\sqrt{2}}(\left|000\right>_{BCD} +\left|010\right>_{BCD}),
\end{equation}
and Z-gate on qubit C, the state $\left|R_1\right>_{BCD}$ is converted to
\begin{equation}
	\left|R_2\right>_{BCD} = \dfrac{1}{\sqrt{2}}(\left|000\right>_{BCD} -\left|010\right>_{BCD}),
\end{equation}
then, CNOT-gate is performed on qubits C and B, taking qubit C as controlling qubit and qubit B as target qubit, the state $\left|R_2\right>_{BCD}$ is converted to
\begin{equation}
	\left|R_3\right>_{BCD} = \dfrac{1}{\sqrt{2}}(\left|000\right>_{BCD} -\left|110\right>_{BCD}),
\end{equation}
and X-gate on qubit B,
\begin{equation}
	\left|R_4\right>_{BCD} = \dfrac{1}{\sqrt{2}}(\left|100\right>_{BCD} -\left|010\right>_{BCD}),
\end{equation}
then, H-gate is performed on qubit B, the state $\left|R_4\right>_{BCD}$ becomes
\begin{equation}
	\left|R_5\right>_{BCD} = \dfrac{1}{2}(\left|000\right>_{BCD} - \left|100\right>_{BCD} - \left|010\right>_{BCD} - \left|110\right>_{BCD}),
\end{equation}
and, performing CNOT-operation on qubits C and D, qubit C works as controlling qubit and D as target qubit, 
\begin{equation}
	\left|R_6\right>_{BCD} = \dfrac{1}{2}(\left|000\right>_{BCD} - \left|100\right>_{BCD} - \left|011\right>_{BCD} - \left|111\right>_{BCD}),
\end{equation}
then, H-gate is performed on qubit D, the state $\left|R_6\right>_{BCD} $ is converted to
\begin{align}
	\left|R_7\right>_{BCD} &= \dfrac{1}{2\sqrt{2}}(\left|000\right>_{BCD} +\left|001\right>_{BCD}  - \left|100\right>_{BCD} - \left|101\right>_{BCD} \nonumber\\
	&	- \left|010\right>_{BCD} +  \left|011\right>_{BCD} - \left|110\right>_{BCD} + \left|111\right>_{BCD}),
\end{align}
after that $	\left|R\right>_{BCD}$ is obtained, by performing a final Z-gate on qubit B (\textbf{Fig.\ref{Fig6}}).
\subsection{Construction of the target state}
\begin{figure}[h!]
	\centering
	\includegraphics[scale=0.3]{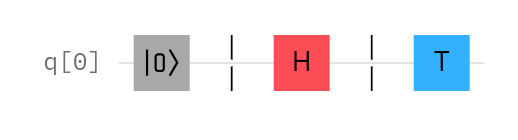}
	\caption{Quantum circuit illustrating the construction of the target state. The coherent state discussed in this figure is a coherent state of a varied phase $\phi = \pi/4$.}\label{Fig7}
\end{figure}

The target state in our protocol, is a coherent state of a varied phase $\phi$ taking the form of Eq.(14), which can be reconstructed by applying H-gate and a phase-gate. The input state will be represented on the quantum circuit by a single quantum qubit initialized to $\left|0\right>$,
\begin{equation}
	\left|in_0\right>_A = \left|0\right>_A,
\end{equation}
then, performing H-gate on qubit A, the initial target state $\left|in_0\right>_A $ is converted to
\begin{equation}
	\left|in_1\right>_A = \dfrac{1}{\sqrt{2}}(\left|0\right>_A + \left|1\right>_A),
\end{equation}
after that, $\left|in\right>_A$ is reconstructed (\textbf{Fig.\ref{Fig7}}), by performing phase-gates ($S^\dagger$-gate, $T^\dagger$-gate, $T$-gate, $S$-gate).
\subsection{Combined state and deferred measurement}
\begin{figure}[h!]
	\centering
	\includegraphics[scale=0.38]{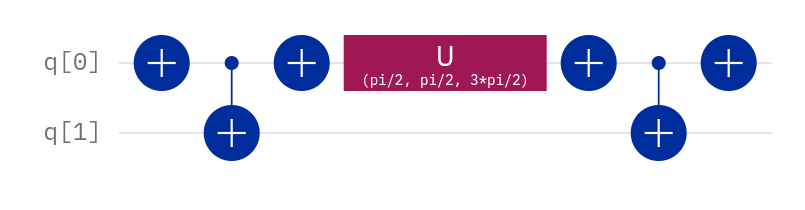}
	\caption{Quantum circuit illustrating the beam splitter used in this scheme.}\label{Fig8}
\end{figure}

The combined state of the target state and the hybrid three-qubit quantum channel, after performing phase-gates on the three qubits of the quantum channel, is transformed to take the forme of Eq.(25). Alice applies CNOT-operation on qubits A and B, where qubit A works as controlling qubit and qubit B as target qubit. Then, Alice performs Hadamard-gate on qubit A (\textbf{Fig.\ref{Fig8}}).
\begin{figure}
	\centering
	\includegraphics[scale=0.25]{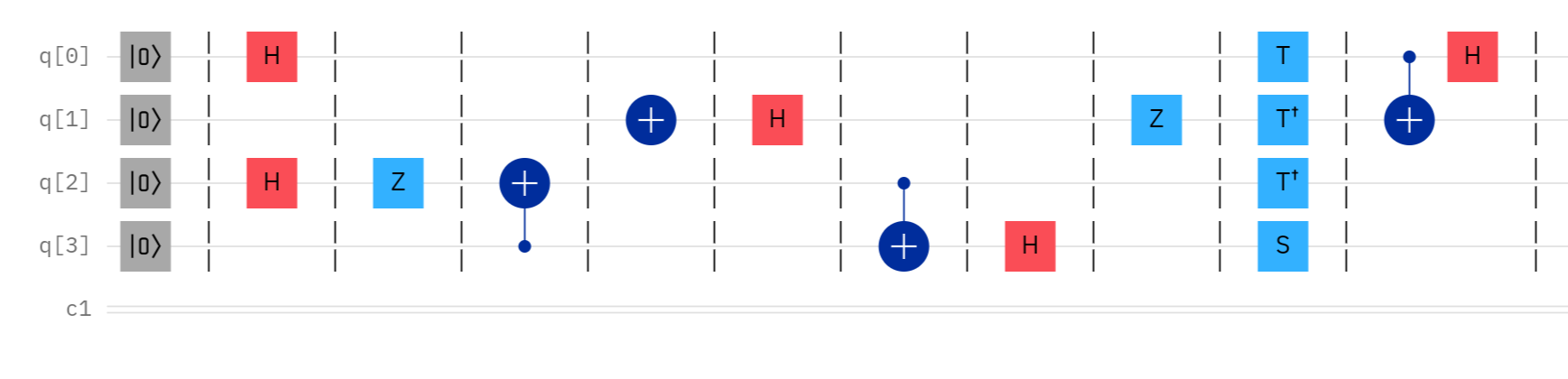}
\caption{Quantum circuit illustrating the combined state of the four subsystems, i.e. input's qubit A, Alice's qubit B, Bob's qubit C and Charlie's qubit D respectively. To choose convenable phases, we relied on the theoretical founded critereon result ($\theta_B=\theta_C$ and $\phi-\theta_B=\pi/2$).}\label{Fig9}
\end{figure} 

Next, Alice qubits A and B pass through a symmetrical beam splitter, by performing CNOT-gates and X-gates (\textbf{Fig.\ref{Fig9}}). In the deferred measurement, CNOT-operations and Controlled-Z (CZ-gate) are performed. CNOT-gate is performed on qubits B and C (B works as controlling qubit and C works as target qubit), and qubits C and D (D works as controlling qubit and C works as target qubit). Then, CZ-gate is applied on qubits A and C, and qubits C and D. After that, Alice destroys qubits on its possession and sends the result of her measurement to Bob by classical way. Charlie, also measures his qubit and sends his qubit information to Bob by classical way. After receiving the information from Alice and Charlie, Bob applies appropriate unitary Pauli operators and measures his qubit. The experimental below is performed on "IBM quantum features", where stastical data comparison of the teleportation of a coherent state of a varied phase $\phi$ between "IBM-qasm simulator", "IBM-aer simulator", "IBM-belem simulator" is presented.
\begin{figure}[h!]
	\centering
	\includegraphics[scale=0.28]{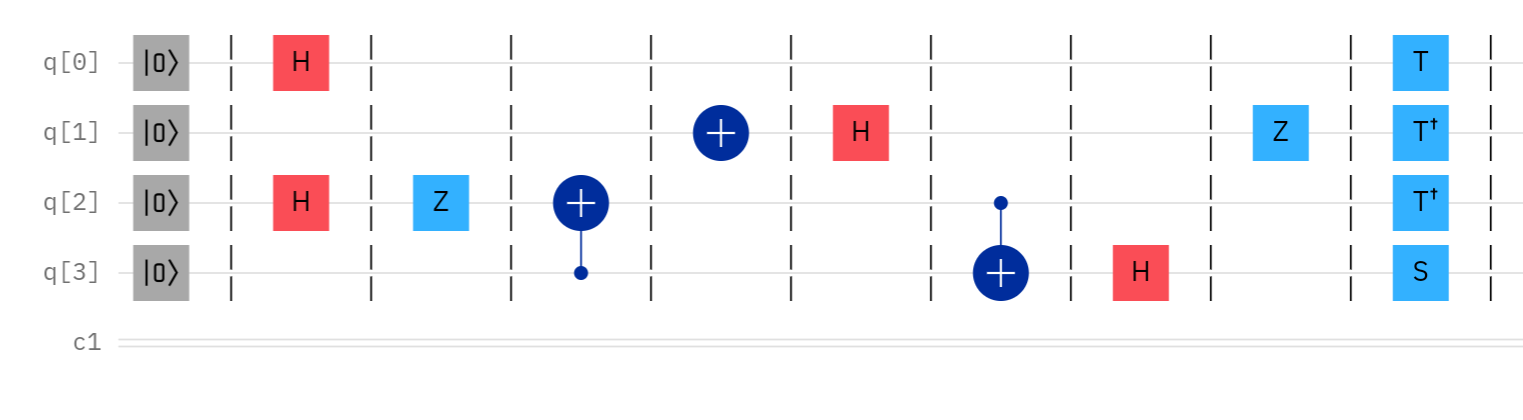}
	\includegraphics[scale=0.3]{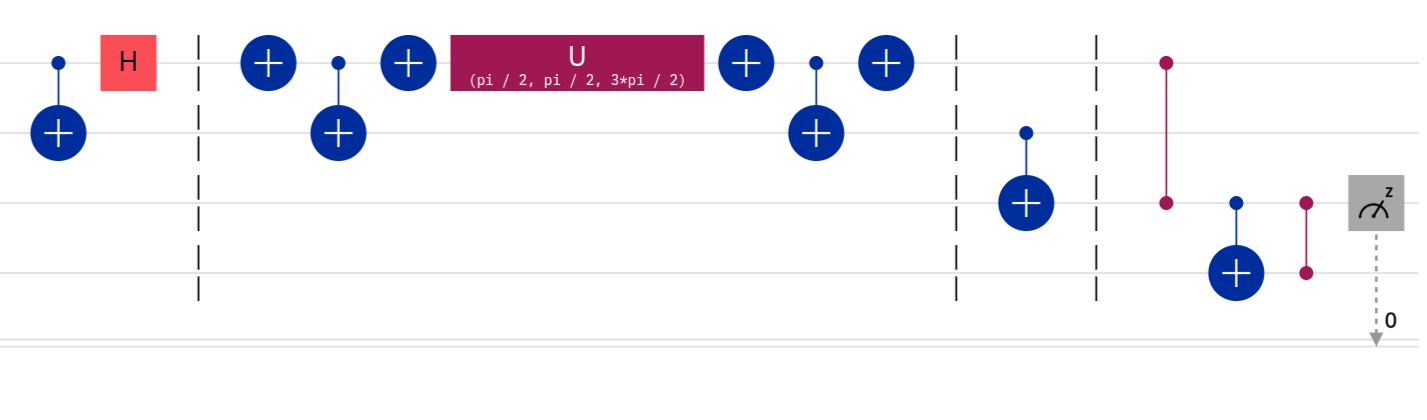}
	\caption{Quantum circuit for CQT of a coherent state of a varied phase (here, we take for example $\phi=\pi/4$) using an hybrid tripartite quantum channel.}\label{Fig10}
\end{figure}
	
\begin{center}
	\begin{table}[h!]
		\centering
	\renewcommand{\arraystretch}{1.3}
	\begin{tabular}{|p{1.5cm}|p{2cm}|p{1.9cm}|p{1.5cm}|p{2.5cm}|p{2.5cm}|p{2.5cm}|}
		\hline \rowcolor{lightgray} Input's phase  & Input's probability& Alice's and Bob's phase& Theor.  output & IBM-qasm simulator& IBM-Aer simulator  &  IBMQ-Belem simulator  \\
		\hline  $0 $  & $[0.5;0.5]$  & $-\pi/2$ & $[0.5;0.5]$ &$[0.502;0.498]$ &$[0.499;0.501]$ & $[0.520;0.480]$ \\
		\hline  $\pi/4$  & $[0.5;0.5]$  & $-\pi/4$ & $[0.5;0.5]$& $[0.500;0.500]$ & $[0.498;0.502]$ & $[0.528;0.472]$ \\
		\hline  $\pi/2$  & $[0.5;0.5]$  & $0$ & $[0.5;0.5]$&$[0.493;0.507]$&$[0.499;0.501]$& $[0.553;0.447]$ \\
		\hline  $3\pi/4$  &$[0.5;0.5]$  & $\pi/4$ & $[0.5;0.5]$&$[0.499;0.501]$ & $[0.508;0.492]$&$[0.537;0.463]$  \\
		\hline  $\pi $  & $[0.5;0.5]$  & $\pi/2$ & $[0.5;0.5]$& $[0.490;0.510]$ & $[0.496;0.504]$ & $[0.524;0.476]$ \\
		\hline  $5\pi/4 $  & $[0.5;0.5]$  & $3\pi/4$ & $[0.5;0.5]$&$[0.501;0.499]$ &$[0.498;0.502]$ & $[0.544;0.456]$ \\
		\hline  $3\pi/2$  & $[0.5;0.5]$  & $\pi$ & $[0.5;0.5]$&$[0.495;0.505]$ &$[0.501;0.499]$ & $[0.516;0.484]$ \\
		\hline  $7\pi/4$  & $[0.5;0.5]$  & $5\pi/4$ & $[0.5;0.5]$ &$[0.497;0.503]$ &$[0.496;0.504]$ & $[0.519;0.480]$ \\
		\hline
	\end{tabular}
	\caption{Comparison between theoretical and experimental results. Probabilities for CQT of a coherent state of a varied phase obtained by "IBM-Qasm simulator", "IBM-Aer simulator" and "IBMQ-Belem simulator" (with 8192 shots to increase result's accuracy and to reduce statistical errors to minimum). The abreviation $[x;y]$ is used to note output's qubit probability, where $x$ presents the probability to obtain $\left|0\right>$ and $y$ works as the probability to obtain $\left|1\right>$.}
	\label{tab3}
\end{table}
\end{center}
\begin{figure}[h!]
	{{\begin{minipage}[b]{.3\linewidth}
				\centering
				\includegraphics[scale=0.45]{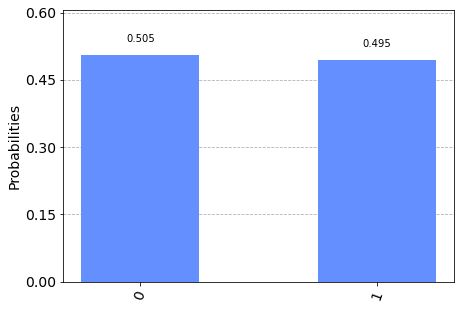}\vfill
				$(a)$
			\end{minipage}\hfill
			\begin{minipage}[b]{.24\linewidth}
				\centering
				\includegraphics[scale=0.45]{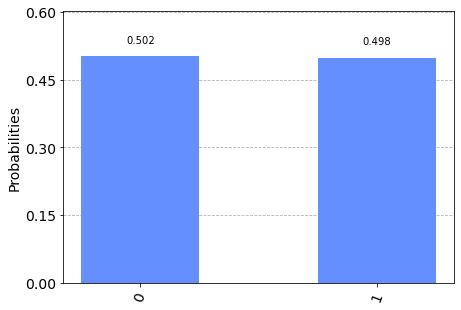}\vfill
				$(b)$
			\end{minipage}\hfill
			\begin{minipage}[b]{.31\linewidth}
				\centering
				\includegraphics[scale=0.45]{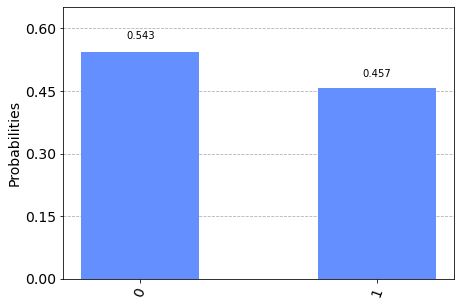}\vfill
				$(c)$
	\end{minipage}}}
	\caption{Histogram: (a) shows the output probability obtained by "IBM-Qasm simlator"; (b) by "IBM-Aer simulator"; and (c) by "IBMQ-Belem simulator" (with 1024 shots).}
\end{figure}
\section{Concluding remarks}\label{IV}
In most controlled quantum teleportation protocols, the sender Alice sends an unknown arbitrary discrete state to the remote receiver Bob, under the control of a third party Charlie, acting as the discrete quantum channel controller. Then unitary Pauli operators are used to completing the transfer protocol. It should be noted that the receiver cannot reconstruct the original arbitrary discrete state without the authorization of the data controller. In this work, we theoretically realize a controlled quantum teleportation protocol using a combination of discrete-valued states and continuous-valued states as an entangled quantum channel pre-shared between the three protocol participants.\par

Specifically, we have proposed a new protocol of controlled quantum teleportation of a CV qubit, where we used a three-mode hybrid quantum resource channel that consists of two discrete modes and one continuous mode. The continuous mode (coherent state) belongs to the sender's particle Alice while the two discrete modes belong respectively to the receiver's part Bob and the supervisor's part Charlie. The CV qubit is at the disposal of the sender. To accomplish this protocol, we replaced the CV qubit by a coherent state of a varied phase which is the decomposition up to the single-photon term in the Fock basis. We discussed the fidelity of the controlled quantum teleportation both in the first and second order of $\alpha$ and we presented their variations. In comparison with the results obtained in Ref.\cite{Ulanov2017} for standard quantum teleportation, our work present more interesting results for controlled quantum teleportation of a coherent state via a tripartite hybrid resource, we found that the perfect teleportation fidelity depends on the phase difference between the phase of the state to teleport and the phase of the sender's mode, we show that for a difference which approaches $0$ or $\pi$, the fidelity reaches its maximum value. Furthermore, we have shown that at second order $\alpha$, the teleportation efficiency remains independent of the amplitude $\alpha$ and the squeezing parameter $\zeta$.\par

Besides, it is worth to noting that running quantum communications using controlled quantum teleportation protocols is absolutely a good choice and a very secure carrier because it cannot be cracked or listened. In this respect, the quantum channel chosen here can be used in ideal conditions in complete security and confidentiality. Finally, the astonishing results obtained in this paper have been theoretically and experimentally demonstrated their effectiveness in the transmission of information.

\appendix
\section{Measurements on continuous variables}
Similar to quantum states, quantum measurements on continuous states can be divided into two broad categories: Gaussian quantum measurements and non-Gaussian quantum measurements. The first category contains two types of Gaussian measurements; the homodyne detection measurement and the heterodyne detection measurement. In the non-Gaussian category, we find the photon counting (PC). The table below summarizes these measures \cite{Yuen1978,Damiani1997}.
\begin{table}[h!]
	\centering
	\renewcommand{\arraystretch}{1.3}
	\begin{tabular}{|p{1.9cm}||p{2.2cm}||p{12cm}|}
		\hline \rowcolor{lightgray} Category & Measure & Definition\\
		\hline Gaussian & Homodyne detection & Homodyne detection is a projective measurement on the eigenstates $\left|x \right\rangle_{\phi}$ of the quadrature operator $\widehat{q}_{\phi}$. In this case, $\widehat{q}_{\phi} = \widehat{q}\cos\phi + \widehat{p}\sin\phi$ is the projective operator.\\
		\cline{2-3} & Heterodyne detection & The heterodyne detection can be viewed as a simultaneous measurement of $\widehat{q}$ and $\widehat{p}$. Since $\widehat{q}$ and $\widehat{p}$ do not commute, we can consider it as a projection onto coherent state with $\left|n\right>\left<n\right|/\pi$ is the projection operator.\\
		\hline Non-Gaussian & Photon counters &  Use only with a system adopting Fock representation with the projection operator $\left|n\right>\left<n\right|$.\\
		\hline
	\end{tabular}		
	\caption{Summary of the usual quantum measurement in continuous systems.}
\end{table}
\section{Comparing squeezed vacuum and cat state (Even and Odd)}
By analogy to the displacement operator $D\left(\alpha\right)$ \cite{Agrawal1991} that serves to find a coherent state from the vacuum quantum state $\left|0\right>$, the squeezing operator $\widehat{S}\left(\zeta\right)$ is defined as	
\begin{equation}
	\widehat{S}(\zeta) = \exp\left[\dfrac{1}{2}\left(\zeta^* a^2 - \zeta a^{\dagger 2}\right)\right],
\end{equation}
where $a$ and $a^\dagger$ are the annihilation and creation operators, and $\zeta$ is the squeezing factor \cite{Walles1983}. The most used squeezed state is the squeezed vacuum, which results from the parametric amplification of the quantum vacuum $\left|0\right>$ as
\begin{align}
	\left	|0,\zeta\right> = \widehat{S}(\zeta)\left|0\right>
	= \dfrac{1}{\sqrt{\cosh \zeta}} \sum_{n=0}^{\infty}\dfrac{\sqrt{(2n)!}}{2^{n} n!} \tanh \zeta^n \left|2n\right>,\label{eqB2}
\end{align}
and can be decomposed in the Fock basis as	
\begin{equation}
	\left|0, \zeta \right>=\left|0\right> +\dfrac{\zeta}{\sqrt{2}}\left|2\right> + \sqrt{\dfrac{3}{8}}\zeta^2\left|4\right>+ O(\zeta^3).\label{eqB3}
\end{equation}
If we apply the photon annihilation operator $\widehat{a}$ on equation (\ref{eqB2}), we obtain
\begin{equation}
	\widehat{a}\widehat{S}(\zeta)\left|0\right>=\zeta\left[\left|1\right> + \sqrt{\dfrac{3}{2}}\zeta\left|3\right>+ O(\zeta^2)\right].\label{eqB4}
\end{equation}
By applying the Fock representation to a coherent state, we can decompose the even and odd coherent states $\left|ODD,\alpha\right>$ and $\left|EVEN,\alpha \right>$ as follows
\begin{equation}
	\left|EVEN, \alpha\right >= \left|0\right> +\dfrac{\alpha^2}{\sqrt{2}}\left|2\right> + O(\alpha^4),
\end{equation}
and
\begin{equation}
	\left	|ODD, \alpha\right >= \left|1\right> +\dfrac{\alpha^2}{\sqrt{6}}\left|3\right> + O(\alpha^4).
\end{equation}
By comparing Eq.(\ref{eqB2}) by Eq.(\ref{eqB3}), we see that if $\zeta<1$, a squeezed vacuum state approximates a even state $\left|EVEN, \alpha \right>$  with $\alpha_+=\sqrt{\zeta}$. In the same way, we comparing Eq.(\ref{eqB2}) and Eq.(\ref{eqB4}), we found that also when $\zeta<1$, a photon-annihilated squeezed vacuum state is approximately an odd state $\left|ODD, \alpha\right >$  with $\alpha_{-}=\sqrt{3\zeta}$. If we take $\zeta=0.18$ as in Ref.\cite{Ulanov2017}, we find the theoretical values of $\alpha_{+}=0.42$ and $\alpha_{-}=0.73$.

\begin{acknowledgements}
AS and NI would like to thank the Abdus Salam International Centre for Theoretical Physics (ICTP-Trieste) for their hospitality during their visit, where some of the work was carried out.
\end{acknowledgements}

\end{document}